\documentclass[aps,pra,10pt,twocolumn,groupedaddress,showpacs,floatfix]{revtex4-1}
\usepackage{dcolumn}
\usepackage{amsmath}
\usepackage{bm}
\usepackage{times}
\usepackage{graphicx}
\usepackage{xcolor}
\newcolumntype{.}{D{x}{}{-1}}
\newcommand*{\cent}[1]{\multicolumn{1}{c}{$#1$}}
\newcommand*{\centt}[1]{\multicolumn{1}{c}{#1}}
\newcolumntype{w}[1]{D{.}{.}{#1}}

\newcommand{\plus}{\phantom{-}}
\newcommand{\icm}{\mathrm{cm}^{-1}}
\newcommand{\rECG}{{\em r}ECG\,}

\begin{document}
\preprint{Version 1.0}

\title{Relativistic corrections for the ground electronic state of molecular hydrogen}
\author{Mariusz Puchalski} 
\affiliation{Faculty of Chemistry, Adam Mickiewicz University,
             Umultowska 89b, 61-614 Pozna{\'n}, Poland}
\author{Jacek Komasa}  
\affiliation{Faculty of Chemistry, Adam Mickiewicz University,
             Umultowska 89b, 61-614 Pozna{\'n}, Poland}
\author{Krzysztof Pachucki}
\affiliation{Faculty of Physics, University of Warsaw, Pasteura 5, 02-093 Warsaw, Poland}

\date{\today}

\begin{abstract}
We recalculate the leading relativistic corrections for the ground electronic state of the hydrogen molecule 
using variational method with explicitly correlated functions which satisfy the interelectronic cusp condition. 
The new computational approach allowed for the control of the numerical precision which reached 
about 8 significant digits. More importantly, the updated theoretical energies became 
discrepant with the known experimental values and we conclude that the yet unknown relativistic 
recoil corrections might be larger than previously anticipated. 
\end{abstract}

\maketitle

\section{Introduction}
Theoretical studies of hydrogen molecule is the cornerstone of the molecular quantum mechanics.
Due to its simplicity, the achieved precision is the highest 
among all molecules and still has a potential of significant enhancement. 
This high precision of theoretical predictions for H$_2$ leads to improved tests of quantum 
electrodynamics and improved bounds on hypothetical interactions \cite{ubachs}.
Moreover, at the $10^{-7}$ cm$^{-1}$ precision level the dissociation energy is sensitive to
the proton charge radius, which may help to resolve the so called proton radius conundrum \cite{Pohl:10}.
This requires high accuracy calculations of not only nonrelativistic energies, but also leading relativistic 
$O(\alpha^2)$, QED $O(\alpha^3)$, as well as the higher order corrections $O(\alpha^4)$ and  $O(\alpha^5)$.
In fact, the nonrelativistic energies can already be calculated with the precision of $10^{-7}$ cm$^{-1}$, 
as demonstrated in Ref.~\cite{PK16}. The $O(\alpha^4)$ contribution has very recently been 
calculated \cite{PKCP16} using explicitly correlated Gaussian (ECG) functions with $1+r_{12}/2$ prefactor (\rECG) 
that makes the interelectronic cusp condition to be exactly satisfied.
Here, we report the results for the leading $O(\alpha^2)$ relativistic correction using \rECG functions 
and conclude that the compilation of previous results in Ref.~\cite{PLPKPJ09} has  
underestimated numerical uncertainties. We improve the numerical precision by 3-4 orders of magnitude 
and present in detail our computational approach.

\section{Computational method}

In the Born-Oppenheimer (BO) approximation the total wave function is assumed to be a product 
of the electronic and nuclear functions. The Schr{\"o}dinger equation for the electronic wave functions
in the infinite nuclear mass limit (assuming atomic units) is 
\begin{equation}\label{SE}
H \psi(\vec r_1, \vec r_2) = {\cal E}(R) \psi(\vec r_1, \vec r_2)
\end{equation}
where
\begin{equation} \label{Ham}
H = \frac{1}{2} \big( \vec p_1^{\,2} + \vec p_2^{\,2}\big) + V
\end{equation}
and
\begin{equation}
V = \frac{1}{R} -\frac{1}{r_{1A}} - \frac{1}{r_{2A}}
    -\frac{1}{r_{1B}} - \frac{1}{r_{2B}} +\frac{1}{r} 
\end{equation}
with  $R = r_{AB}$ and $r=r_{12}$, and where  indices 1 and 2 correspond to electrons, 
whereas $A$ and $B$ correspond to the nuclei. 
The leading relativistic correction in the BO  approximation, 
is that ${\cal E}_\mathrm{rel}(R)$ to the nonrelativistic potential ${\cal E}(R)$. 
This correction can be expressed in terms of the expectation value
\begin{equation}
{\cal E}_\mathrm{rel}(R) = \langle \psi | H_\mathrm{rel}| \psi \rangle
\label{ERdirect}
\end{equation}
of the Breit-Pauli Hamiltonian \cite{Bethe:57} 
\begin{eqnarray}
H_\mathrm{rel} &=& -\frac{1}{8} \big(p_1^4 + p_2^4\big)
 +\pi\,\delta^3(r)
-\frac{1}{2}\,p_1^i\,
\biggl(\frac{\delta^{ij}}{r}+\frac{r^i\,r^j}{r^3}\biggr)\,p_2^j
\nonumber \\ &&
+\frac{\pi}{2}\big(\,\delta^3(r_{1A}) +\delta^3(r_{2A})
+\delta^3(r_{1B})
+\delta^3(r_{2B}) \big), \label{HBP}
\end{eqnarray}
where we neglected spin dependent terms vanishing for the ground electronic state of $^1\Sigma_g^+$ symmetry.
The accurate calculation of the above expectation value is the principal goal of this work.
We assume that every \rECG basis function contains the $1+r/2$ factor and perform
special transformation (regularization) of matrix elements, including those with
the Dirac-$\delta$ function \cite{Drachman:81}, for improving the numerical convergence.
We demonstrate a significant enhancement in numerical precision
and indicate that previous numerical results \cite{PLPKPJ09} were not as accurate as claimed.
In order to be more convincing, we provide results obtained in three approaches: 
i) direct (no regularization) with ECG, ii) standard regularization with ECG, 
and iii) modified regularization with \rECG functions.  
To test the convergence of these three different approaches, at first we perform calculations for $R=0$, 
namely for the helium atom, for which highly accurate reference results can be obtained using explicitly 
correlated exponential functions. Next, the ECG calculations are performed for molecular hydrogen. 
Comparison of individual operators from different approaches is presented for the equilibrium 
internuclear distance, namely for $R = 1.4$ a.u. The most accurate predictions were obtained for 
the regularization with \rECG functions at 53 points in the range $R=0.0 - 10$ a.u. 

Except for our recent paper \cite{PKCP16}, there has been no similar study of regularization techniques
in literature due to difficulties with two-center integrals
involving inverse powers of interparticle distances. In relation to this,
we have introduced a novel algorithm for numerical quadrature 
of non-standard ECG two-center integrals \cite{PKCP16}, which enables 
very efficient calculations of all complicated matrix elements. 

\section{Regularization of the relativistic correction}

In this section we provide regularization formulas for 
matrix elements with Dirac-$\delta$ and $p^4$ operators in $H_\mathrm{rel}$.
The latter operator can be regularized according to the two schemes: the standard one, 
already employed in the past in quantum molecular computations, \cite{Wolniewicz:93}
and the modified scheme, valid in the case of the wave function obeying 
the Kato's cusp condition.

According to the standard scheme the relativistic operators are transformed into the
regular form by the following relations
\begin{align}
\label{drXa}
4 \pi\, \delta^3(r_{1A}) &= 4 \pi \, [\delta^3(r_{1A})]_r + \bigg\{\frac{2}{r_{1A}} ,H-{\cal E} \bigg\}\,,  \\
\label{drab}
4 \pi\, \delta^3(r) &= 4 \pi \, [\delta^3(r)]_r + \bigg\{\frac{1}{r} , H-{\cal E} \bigg\}\,,  \\
\label{pat4}
p_1^4 + p_2^4  &= [p_1^4 + p_2^4]_r +  4\,\Big\{{\cal E}-V , H-{\cal E} \Big\} + 4\,(H-{\cal E})^2,
\end{align}
where
\begin{align}
4 \pi \, [\delta^3(r_{1A})]_r &= \frac{4}{r_{1A}} ({\cal E}-V) - \vec p_1\,\frac{2}{r_{1A}}\,\vec p_1
- \vec p_2\,\frac{2}{r_{1A}}\,\vec p_2\,,\label{drXar}\\
4 \pi \, [\delta^3(r)]_r &= \frac{2}{r} ({\cal E}-V) - \vec p_1\,\frac{1}{r}\,\vec p_1
- \vec p_2\,\frac{1}{r}\,\vec p_2\,,\label{drabr}\\
[p_1^4 + p_2^4]_r &= 4 \,({\cal E}-V)^2 - 2\,  p_1^{2} \, p_2^{2}\,.
\label{pat4r}
\end{align}
For the exact wave function $\psi$, which fulfills the electronic Schr{\"o}dinger equation
$({\cal E}-H)\psi=0$, the expectation value identity holds 
$\langle \psi | \ldots | \psi \rangle = \langle \psi | [\ldots]_r | \psi \rangle$,
since for an arbitrary operator $Q$, $\langle \psi | \{Q , H-{\cal E}\} | \psi \rangle = 0$. 
For an approximate function $\tilde{\psi}$, such expectation values do not vanish, but converge
to zero in the limit $\tilde{\psi}\to\psi$. In practice, the numerical convergence
of the regularized form is much faster, so the leading relativistic correction
shall be evaluated as
\begin{equation}
{\cal E}_\mathrm{rel}(R) = \langle \psi | [H_\mathrm{rel}]_r | \psi \rangle
\label{HBPrr} 
\end{equation}
with
\begin{eqnarray}
\plus[H_\mathrm{rel}]_r &=& -\frac{1}{8} [p_1^4 + p_2^4]_r
+\frac{\pi}{2}\big(\,[\delta^3(r_{1A})]_r + [\delta^3(r_{2A})]_r \nonumber \\ &&
+ [\delta^3(r_{1B})]_r
+ [\delta^3(r_{2B})]_r \big) 
 +\pi\,[\delta^3(r)]_r 
 \nonumber \\ &&
-\frac{1}{2}\,p_1^i\,
\biggl(\frac{\delta^{ij}}{r}+\frac{r^i\,r^j}{r^3}\biggr)\,p_2^j\,.
\label{Hrelr}
\end{eqnarray}

The modified regularization is applied when the wave function $\tilde{\psi}$ 
exactly satisfies the interelectronic cusp condition, as for example the \rECG function does.
In this case, the action of $[p_1^4 + p_2^4]_r$ on such a function can be represented as
\begin{equation}
[p_1^4 + p_2^4]_r |\tilde{\psi}\rangle = \Bigl\{4 \,({\cal E}-V)^2 - 2\,\tilde{p}_1^{\,2}\,\tilde{p}_2^{\,2} +8\pi\,\delta^3(r) \Bigr\}|\tilde{\psi}\rangle.
\label{pat4rtilde}
\end{equation}
The new term $\tilde{p}_1^{\,2}\,\tilde{p}_2^{\,2}$ (in contrast to $p_1^2\,p_2^2$)
is understood as the differentiation $\nabla_1^2 \, \nabla_2^2$ of $\tilde{\psi}$ 
as a function, that is with the omission of the $\delta^3(r)$-term.
Now, if we are interested in determination of the $\langle\tilde{\psi} | p_1^4 + p_2^4 |\tilde{\psi}\rangle$ alone,
we can additionally replace the Dirac-$\delta$ operator by its regularized form and obtain
the 'fully-regularized' expectation value
\begin{equation}
\langle \psi | p_1^4 + p_2^4 |\psi\rangle = 
\langle \psi | 4 \,({\cal E}-V)^2 - 2\,\tilde{p}_1^{\,2}\,\tilde{p}_2^{\,2} +8\pi\,[\delta^3(r)]_r |\psi\rangle.
\end{equation}
We emphasize that here, unlike $p_1^2\,p_2^2$, the $\tilde{p}_1^{\,2}\,\tilde{p}_2^{\,2}$ term
differentiates the right-hand-side wave function only.
The specific relationship~(\ref{pat4rtilde}) can also be further employed to simplify the expectation value 
of the overall Breit-Pauli Hamiltonian~(\ref{HBP}) by complete elimination of the  $\pi\,\delta^3(r)$ term
\begin{align}
{\cal E}_\mathrm{rel}(R) =& \langle \psi | [H_\mathrm{rel}]'_r | \psi \rangle, \label{HBPr} \\
[H_\mathrm{rel}]'_r =& 
-\frac{1}{2}({\cal E}-V)\left({\cal E}-\frac{1}{R}-\frac{1}{r}\right) \nonumber \\ &
+ \frac{1}{4}\left(\tilde{p}_1^{\,2}\,\tilde{p}_2^{\,2}
+ \vec p_1\,\tilde{V}\,\vec p_1
+ \vec p_2\,\tilde{V}\,\vec p_2 \right) \nonumber\\
& -\frac{1}{2}\,p_1^i\,\biggl(\frac{\delta^{ij}}{r}+\frac{r^i\,r^j}{r^3}\biggr)\,p_2^j \label{HBPp}
\end{align}
with 
$\tilde{V} = -1/r_{1A} - 1/r_{1B} - 1/r_{2A} - 1/r_{2B}$.
Apart from its compactness, this formula has an additional important advantage
which is not readily noticeable. Due to the above cancellations, 
all the time-consuming integrals with three odd powers 
of interparticle distances do not appear in the matrix elements with {\it r}ECG functions. 
This non-trivial cancellation has remarkable impact on calculations of the relativistic correction.

\section{Integrals with ECG functions}
The variational wave function 
\begin{eqnarray}
 \psi &=& \sum_i c_i \psi_i(\vec r_1, \vec r_2)\,, \\
 \psi_i &=& (1 + \hat i)\,(1+P_{1\leftrightarrow 2})\,\phi_i (\vec r_1, \vec r_2),
 \end{eqnarray}
where $\hat{i}$ and $P_{1\leftrightarrow 2}$ are the inversion and the electron exchange operators,
can be accurately represented in the basis of ECG functions of the form 
\begin{equation}
\phi_{\Sigma^+} = e^{-a_{1A}\,r_{1A}^2 -a_{1B}\,r_{1B}^2 -a_{2A}\,r_{2A}^2 -a_{2B}\,r_{2B}^2 - a_{12}\,r^2 } \label{ECG}
\end{equation}
or in the basis of the modified \rECG functions
\begin{equation}
\phi_{\Sigma^+} = \Bigl(1+\frac{r}{2}\Bigr)
\,e^{-a_{1A}\,r_{1A}^2 -a_{1B}\,r_{1B}^2 -a_{2A}\,r_{2A}^2 -a_{2B}\,r_{2B}^2 - a_{12}\,r^2 }\,. \label{r12ECG}
\end{equation}
Nonlinear $a$-parameters are determined variationally for every ECG or \rECG basis function
and linear $c$-parameters come from the solution of the general eigenvalue problem. 
The primary advantage of ECG type of functions is that all
integrals necessary for the calculations of nonrelativistic and relativistic operators
can be evaluated very effectively as described below.

Each matrix element can be expressed as a linear combination of the following ECG integrals
\begin{eqnarray}
f(n_1,n_2,n_3,n_4,n_5) &=& \frac{1}{\pi^3}\int d^3 r_1 \int d^3 r_2 \, 
r_{1A}^{n_1} r_{1B}^{n_2} r_{2A}^{n_3} r_{2B}^{n_4} r_{12}^{n_5}
\nonumber \\ && \hspace*{-10ex}
\times e^{-a_{1A}\,r_{1A}^2 -a_{1B}\,r_{1B}^2 -a_{2A}\,r_{2A}^2 -a_{2B}\,r_{2B}^2 - a_{12}\,r_{12}^2 } 
\label{27}
\end{eqnarray}
with integers $n_i$ and real parameters $a$. 
Among all the integrals represented by the above formula we can distinguish
two subsets that can be evaluated analytically. The first subset contains the {\it regular}
ECG integrals with the non-negative even integers $n_i$ such that $\sum_i n_i \leq \Omega_1$,
where the shell parameter $\Omega_1=0,2,4,\dots$. These integrals can be generated 
by differentiation over $a$-parameters of the following master integral
\begin{eqnarray}
f(0,0,0,0,0) = X^{-3/2} e^{-R^2 \frac{Y}{X}}, \label{regular}
\end{eqnarray}
where
\begin{eqnarray}
X &=& (a_{1A} + a_{1B} + a_{12}) (a_{2A} + a_{2B} + a_{12}) - a_{12}^2 \\
Y &=&   (a_{1B} + a_{1A})\, a_{2A} a_{2B} + a_{1A} a_{1B} (a_{2A} + a_{2B}) 
\nonumber \\ &&
+ a_{12} (a_{1A} + a_{2A}) (a_{1B} + a_{2B}) \,.
\end{eqnarray}
Each differentiation raises one of the $n_i$ exponents by two.
The second subset of integrals permits a single odd index $n_i\geq-1$ 
for which $\sum_i n_i \leq \Omega_2$ ($\Omega_2=-1,1,3,\dots$). 
These so called {\it Coulomb} ECG integrals can also be obtained analytically by differentiation
of another master integral. For instance, when $n_1 = -1$ the master integral reads
\begin{equation}
f(-1,0,0,0,0) = \frac{1}{X \sqrt{X_1}}\,e^{-R^2 \frac{Y}{X}}
F \bigg[R^2\bigg( \frac{Y_1}{X_1} - \frac{Y}{X}\bigg) \bigg], \label{Coulomb}
\end{equation}
where $X_1 = \partial_{a_{1A}} X$, $Y_1 = \partial_{a_{1A}} Y$, and $F(x) = {\rm erf}(x)/x$.

In the standard use of ECG functions the {\it regular} integrals with $\Omega_1=2$ 
and {\it Coulomb} with $\Omega_2=-1$ are sufficient to evaluate matrix elements 
of the electronic Schr{\"o}dinger equation~(\ref{SE}) and thus to perform
calculations of the nonrelativistic energy of the ground state in molecular hydrogen. 
If additionally the analytic gradient minimization is employed, the integrals with 
$\Omega_1=4$ and $\Omega_2=1$ are required. Such nonrelativistic calculations 
have been widely used for many atomic and molecular systems.\citep{Mitroy:13}

The molecular ECG integrals, as opposed to the atomic ones, have no known analytic form 
when two or more $n_i$-s are odd. Such {\em extended} integrals originate from regularization of
the relativistic operators, for example from $V^2$ in Eq.~(\ref{pat4r}),
or from matrix elements of the nonrelativistic Hamiltonian with $r$ECG basis.
The algorithm for numerical evaluation of this {\em extended} type of integrals
relies on the following relation, which decreases one of the indices by one
\begin{eqnarray}\label{gext}
f(n_1-1,n_2,n_3,n_4,n_5) &=&  \\ 
&& \hspace{-2.5cm} \frac{2}{\sqrt{\pi}} 
\int_{0}^\infty {\rm d} y\,f(n_1,n_2,n_3,n_4,n_5)\biggr|_{a_{1A}\rightarrow a_{1A} + y^2} \,.\nonumber 
\end{eqnarray}
The right-hand-side  $f$ is understood as the integral $f(n_1,n_2,n_3,n_4,n_5)$ 
evaluated with the $a_{1A}$ parameter replaced by $a_{1A} + y^2$.
The transformation $y = -1 + 1/x$ converts the infinite integration domain to the finite interval $(0,1)$
for which an $m$-point generalized Gaussian quadrature with logarithmic end-point singularity \cite{Pachucki:14}
is applied
\begin{align} 
&\int_0^1 dx \bigl[W_1(x) + \ln(x)\,W_2(x)\bigr]\nonumber \\
&=\sum_{i=1}^m\,w_i\,\bigl[W_1(x_i) + \ln(x_i)\,W_2(x_i)\bigr]\,.
\label{dfrokhlin}
\end{align}
The $W_{1,2}$ are arbitrary polynomials of maximal degree $m-1$, $w_i$ are weights, 
and $x_i$ are nodes. In terms of this quadrature the integral~(\ref{gext})
can be approximated by the formula
\begin{eqnarray}
f(n_1-1,n_2,n_3,n_4,n_5) &=&  \frac{2}{\sqrt{\pi}} \sum_{i=1}^m \, w_i \, (y_i+1)^{2} \\ 
&& \hspace{-2cm} \times f(n_1,n_2,n_3,n_4,n_5)\biggr|_{a_{1A}\rightarrow a_{1A} + y_i^2}\,. \nonumber 
\label{dfnumgm}
\end{eqnarray}
This quadrature is very efficient for {\em extended} integrals with two odd indices, for which typically
only $m=30$ nodes allows about 16 significant digits to be obtained.  These extended integrals are 
sufficient not only for all the relativistic operators with the regularization applied to ECG wave function,
but also for the modified regularization~(\ref{HBPr})-(\ref{HBPp}) of $H_\mathrm{rel}$ with $r$ECG wave function.

Nevertheless, calculations with \rECG wave function of the expectation values 
of the individual relativistic operators
involve {\em extended} integrals with three odd indices. They can be obtained by the double numerical integration
of {\it Coulomb} ECG integrals over $30^2$ nodes to achieve numerical precision of about 16 significant digits.
This two dimensional integration is numerically stable, but time consuming.

\section{Calculations of relativistic corrections}
Relativistic corrections to the BO potential were calculated according to 
Eqs.~(\ref{ERdirect}), (\ref{HBPrr}), and (\ref{HBPr}).  
In order to demonstrate the convergence of these three different approaches with ECG functions
we compared results at $R=0$, i.e. for the helium atom, to the results obtained with explicitly correlated 
exponential (ECE) functions. Calculations with ECE functions are well known in literature 
(see e.g. Refs.~\cite{Korobov:02} and \cite{Drake:02})
and may serve as an excellent reference point and a rigorous test of the convergence of ECG results.  
Numerical values presented in Table~\ref{TBL-helium} were obtained with 128, 256, 512, and 1024 ECG basis functions.
Direct and standard regularization methods were used with ECG functions
whereas the modified regularization with \rECG functions. We observe a significant enhancement of numerical convergence
of relativistic operators obtained with \rECG basis. The total relativistic correction with $N=1024$
is accurate to 9 digits in the \rECG basis, and to 5-6 digits in the ECG basis.
A similar enhancement is observed for H$_2$ at $R=1.4$ in Table \ref{TBL3}, where we compared our results 
for Dirac-$\delta$ functions with those obtained in the ECE basis. The accuracy of the extrapolated 
value for the total relativistic correction is estimated to have at least 8 significant digits after the decimal dot.
In Table \ref{TBL4} we provide results for the nonrelativistic energy $\mathcal{E}$, 
for the relativistic correction $\mathcal{E}_\mathrm{rel}$, 
and also for all the four individual components of the relativistic correction evaluated at $R\in(0,10)$ a.u. with
the 1024-term basis of \rECG functions. Our results for the overall relativistic correction 
$\mathcal{E}_\mathrm{rel}$  for H$_2$ are estimated to have 8 significant decimal digits.
 
\section{Vibrational averaging}

In order to obtain the final value of the $\alpha^2$ relativistic component of the dissociation
energy, we solved the radial Schr{\"o}dinger equation for two potentials. 
The first potential, used as a reference, is the nonrelativistic (BO) potential 
$\mathcal{E}(R)$ \cite{Pachucki:10b}, which yielded the nonrelativistic (BO) energy level $E$. 
The second was the potential augmented by the relativistic correction $\mathcal{E}_\mathrm{rel}(R)$, 
which gave the eigenvalue corresponding to the relativistic energy level $E + E^{(2)}$. 
The difference between both the eigenvalues $E^{(2)}$ is the relativistic correction to molecular levels.

To establish reliable uncertainties for the final results we studied two sources of error:
the convergence of single point calculations and the polynomial interpolation.
As mentioned above, the relativistic correction was evaluated using basis sets of increasing size,
which permitted a detailed analysis of the convergence at each internuclear distance 
(see Tab.~\ref{TBL3}). From this analysis we estimated that in the vicinity of the equilibrium 
distance the $\mathcal{E}_\mathrm{rel}(R)$ bears an uncertainty of $7\cdot 10^{-9}$ a.u.
equivalent to $2\cdot 10^{-8}\,\icm$.
The influence of the density of the points, at which $\mathcal{E}_\mathrm{rel}(R)$ was evaluated, on
the accuracy of the final result was assessed by doubling the number of points, 
which however were calculated only with 512-term basis. 
As a consequence $E^{(2)}$ was shifted by $5\cdot 10^{-7}\,\icm$.
The related uncertainty due to the selection of the degree of the interpolation polynomial was also 
investigated. By changing the degree in the range $5-12$, we observed changes
in the relativistic correction at the level of $10^{-7}\,\icm$.
To summarize, the largest contribution to the uncertainty of the relativistic 
correction $E^{(2)}$ comes from the limited number of points (and the necessity of interpolation) 
at which the relativistic potential was evaluated. The final relative uncertainty is assumed 
to be smaller than $10^{-6}$ cm$^{-1}$.

\section{Results and summary}

Results of our calculations for the dissociation energy and the two selected most accurately measured 
transitions in H$_2$ and D$_2$ are presented in Tables \ref{TH2BO} and \ref{TD2BO}. 
The nonrelativistic energy $E$ for H$_2$ was calculated by solving the 
full nonadiabatic Schr\"odinger equation in the exponential basis \citep{PK16}, whereas for D$_2$
using the NAPT\cite{PK15} expansion with the neglect of $O(1/\mu)^3$ terms. All the corrections were obtained
within the adiabatic approximation. The relativistic correction $E^{(2)}$ was evaluated
and reported in this work.
The leading QED correction $E^{(3)}$ was obtained in Ref.~\cite{Komasa:11}, 
while the higher order QED, namely $E^{(4)}$ in Ref.~\cite{PKCP16}.
$E^{(5)}$ was estimated from the correction analogous to that of atomic hydrogen with the assumption 
that it is proportional to the electron-nucleus Dirac-$\delta$ and the related uncertainty was 
assigned to be 50\%. 

Our results for the leading relativistic corrections significantly differ from those 
by Piszczatowski {\em et al.} \cite{PLPKPJ09}, whose compilation partially relies on 
the former calculations by Wolniewicz \cite{Wolniewicz:93}.
For example, our relativistic correction to $D_0$ of H$_2$
is $-0.533\,121(1)\,\icm$ whereas Piszczatowski et al. reported
$-0.531\,9(3)\,\icm$. Interestingly, our result is closer
to that by Wolniewicz $-0.533\,0\,\icm$, despite
the differences at the level of individual operators.
In our opinion, these differences come from the much more accurate calculation
of relativistic matrix elements performed here. 

Most importantly, our final theoretical predictions for $D_0$ are now 
in disagreement with experimental values, in contrast to the previous theoretical results \cite{PLPKPJ09}. 
This disagreement, most probably, comes from the underestimation 
of the relativistic nuclear recoil correction. We have previously assumed that these corrections are
of the order of the ratio of the electron mass to the reduced mass of the nuclei, which for H$_2$ is $\sim 10^{-3}$. 
This might be incorrect, because the nonrecoil relativistic correction is anomalously small. 
This assertion is supported by the example of helium atom.
The nonrecoil relativistic correction to the $^4$He ionization energy is $16\,904.024$ MHz,
while the nuclear recoil is $-103.724$ MHz, so the ratio is $6\cdot 10^{-3}$ that is 
an order of magnitude higher than the estimate based on the helium mass ratio $10^{-4}$. 
On the other hand, in the separated atoms limit
the relativistic recoil correction exactly vanishes.
Therefore at present this correction cannot be reliably estimated.

If the relativistic nuclear recoil correction in H$_2$ is underestimated, 
the difference between our predictions and the experimental values for D$_2$ should be smaller than that for H$_2$ 
and this is really the case ({\em vide} Tables \ref{TH2BO} and \ref{TD2BO}). 
We emphasize that our theoretical predictions should be treated as preliminary
until the relativistic nuclear recoil corrections are reliably calculated. 
In fact, such corrections have already been obtained by Stanke and Adamowicz\citep{Stanke:13} 
for purely vibrational states. However, their result for 
the total relativistic dissociation energy of H$_2$ (with $E^{(2)} = -0.5691\,\icm$), when augmented by missing
higher order corrections, yields $D_0=36\,118.0318\,\icm$, which differs from the experimental value 
by as much as $0.038\,\icm$, so its numerical uncertainty is out of control. 
We plan to calculate these nonadiabatic corrections using the 
fully nonadiabatic wave function in exponential basis as in Ref.~\cite{PK16} 
or by using nonadiabatic perturbation theory (NAPT) \cite{PK09}. Certainly,
this calculation has to be performed for resolving discrepancies with  H$_2$ experiments.

In conclusion, the former excellent agreement of theoretical predictions with experimental $D_0$ values 
was accidental and the improved calculations of the leading relativistic corrections result in 
a few $\sigma$ disagreements with experimental values for dissociation energies and transition energies, 
which most probably is caused by the unknown relativistic nuclear recoil (nonadiabatic) effects 
in the relativistic corrections.

\begin{acknowledgments}
We wish to thank Bogumi\l\ Jeziorski for comments to the manuscript. 
This work was supported by the National Science Center (Poland) Grant
Nos. 2012/04/A/ST2/00105 (K.P.) and 2014/13/B/ST4/04598 (M.P. and J.K.), 
as well as by a computing grant from the Poznan Supercomputing and Networking Center, 
and by PL-Grid Infrastructure.
\end{acknowledgments}


\begin{thebibliography}{9}
\expandafter\ifx\csname natexlab\endcsname\relax\def\natexlab#1{#1}\fi
\expandafter\ifx\csname bibnamefont\endcsname\relax
  \def\bibnamefont#1{#1}\fi
\expandafter\ifx\csname bibfnamefont\endcsname\relax
  \def\bibfnamefont#1{#1}\fi
\expandafter\ifx\csname citenamefont\endcsname\relax
  \def\citenamefont#1{#1}\fi
\expandafter\ifx\csname url\endcsname\relax
  \def\url#1{\texttt{#1}}\fi
\expandafter\ifx\csname urlprefix\endcsname\relax\def\urlprefix{URL }\fi
\providecommand{\bibinfo}[2]{#2}
\providecommand{\eprint}[2][]{\url{#2}}

\bibitem{ubachs} W. Ubachs, J.C.J. Koelemeij, K.S.E. Eikema, E.J. Salumbides,
J. Mol. Spectr. {\bf 320}, 1 (2016).

\bibitem[{\citenamefont{Pohl et~al.}(2010)\citenamefont{Pohl, Antognini, Nez,
  Amaro, Biraben, Cardoso, Covita, Dax, Dhawan, Fernandes et~al.}}]{Pohl:10}
\bibinfo{author}{\bibfnamefont{R.}~\bibnamefont{Pohl}},
  \bibinfo{author}{\bibfnamefont{A.}~\bibnamefont{Antognini}},
  \bibinfo{author}{\bibfnamefont{F.}~\bibnamefont{Nez}},
  \bibinfo{author}{\bibfnamefont{F.~D.} \bibnamefont{Amaro}},
  \bibinfo{author}{\bibfnamefont{F.}~\bibnamefont{Biraben}},
  \bibinfo{author}{\bibfnamefont{J.~M.~R.} \bibnamefont{Cardoso}},
  \bibinfo{author}{\bibfnamefont{D.~S.} \bibnamefont{Covita}},
  \bibinfo{author}{\bibfnamefont{A.}~\bibnamefont{Dax}},
  \bibinfo{author}{\bibfnamefont{S.}~\bibnamefont{Dhawan}},
  \bibinfo{author}{\bibfnamefont{L.~M.~P.} \bibnamefont{Fernandes}},
  \bibnamefont{et~al.}, \bibinfo{journal}{Nature}
  \textbf{\bibinfo{volume}{466}}, \bibinfo{pages}{213} (\bibinfo{year}{2010}).

\bibitem[{\citenamefont{Pachucki and Komasa}(2016)}]{PK16}
\bibinfo{author}{\bibfnamefont{K.}~\bibnamefont{Pachucki}} \bibnamefont{and}
  \bibinfo{author}{\bibfnamefont{J.}~\bibnamefont{Komasa}},
  \bibinfo{journal}{J. Chem. Phys.} \textbf{\bibinfo{volume}{144}},
  \bibinfo{pages}{164306} (\bibinfo{year}{2016}).

\bibitem[{\citenamefont{Puchalski et~al.}(2016)\citenamefont{Puchalski, Komasa,
  Czachorowski, and Pachucki}}]{PKCP16}
\bibinfo{author}{\bibfnamefont{M.}~\bibnamefont{Puchalski}},
  \bibinfo{author}{\bibfnamefont{J.}~\bibnamefont{Komasa}},
  \bibinfo{author}{\bibfnamefont{P.}~\bibnamefont{Czachorowski}},
  \bibnamefont{and} \bibinfo{author}{\bibfnamefont{K.}~\bibnamefont{Pachucki}},
  \bibinfo{journal}{Phys. Rev. Lett.} \textbf{\bibinfo{volume}{117}},
  \bibinfo{pages}{263002} (\bibinfo{year}{2016}).

\bibitem[{\citenamefont{Piszczatowski et~al.}(2009)\citenamefont{Piszczatowski,
  Lach, Przybytek, Komasa, Pachucki, and Jeziorski}}]{PLPKPJ09}
\bibinfo{author}{\bibfnamefont{K.}~\bibnamefont{Piszczatowski}},
  \bibinfo{author}{\bibfnamefont{G.}~\bibnamefont{Lach}},
  \bibinfo{author}{\bibfnamefont{M.}~\bibnamefont{Przybytek}},
  \bibinfo{author}{\bibfnamefont{J.}~\bibnamefont{Komasa}},
  \bibinfo{author}{\bibfnamefont{K.}~\bibnamefont{Pachucki}}, \bibnamefont{and}
  \bibinfo{author}{\bibfnamefont{B.}~\bibnamefont{Jeziorski}},
  \bibinfo{journal}{J. Chem. Theory Comput.} \textbf{\bibinfo{volume}{5}},
  \bibinfo{pages}{3039} (\bibinfo{year}{2009}).

\bibitem[{\citenamefont{Bethe and Salpeter}(1957)}]{Bethe:57}
\bibinfo{author}{\bibfnamefont{H.~A.} \bibnamefont{Bethe}} \bibnamefont{and}
  \bibinfo{author}{\bibfnamefont{E.~E.} \bibnamefont{Salpeter}},
  \emph{\bibinfo{title}{Quantum Mechanics of One- and Two-Electron Systems}}
  (\bibinfo{publisher}{Springer-Verlag, Berlin and New York},
  \bibinfo{year}{1957}).

\bibitem[{\citenamefont{Drachman}(1981)}]{Drachman:81}
\bibinfo{author}{\bibfnamefont{R.~J.} \bibnamefont{Drachman}},
  \bibinfo{journal}{J. Phys. B} \textbf{\bibinfo{volume}{14}},
  \bibinfo{pages}{2733} (\bibinfo{year}{1981}).

\bibitem[{\citenamefont{Wolniewicz}(1993)}]{Wolniewicz:93}
\bibinfo{author}{\bibfnamefont{L.}~\bibnamefont{Wolniewicz}},
  \bibinfo{journal}{J. Chem. Phys.} \textbf{\bibinfo{volume}{99}},
  \bibinfo{pages}{1851} (\bibinfo{year}{1993}).

\bibitem[{\citenamefont{Mitroy et~al.}(2013)\citenamefont{Mitroy, Bubin,
  Horiuchi, Suzuki, Adamowicz, Cencek, Szalewicz, Komasa, Blume, and
  Varga}}]{Mitroy:13}
\bibinfo{author}{\bibfnamefont{J.}~\bibnamefont{Mitroy}},
  \bibinfo{author}{\bibfnamefont{S.}~\bibnamefont{Bubin}},
  \bibinfo{author}{\bibfnamefont{W.}~\bibnamefont{Horiuchi}},
  \bibinfo{author}{\bibfnamefont{Y.}~\bibnamefont{Suzuki}},
  \bibinfo{author}{\bibfnamefont{L.}~\bibnamefont{Adamowicz}},
  \bibinfo{author}{\bibfnamefont{W.}~\bibnamefont{Cencek}},
  \bibinfo{author}{\bibfnamefont{K.}~\bibnamefont{Szalewicz}},
  \bibinfo{author}{\bibfnamefont{J.}~\bibnamefont{Komasa}},
  \bibinfo{author}{\bibfnamefont{D.}~\bibnamefont{Blume}}, \bibnamefont{and}
  \bibinfo{author}{\bibfnamefont{K.}~\bibnamefont{Varga}},
  \bibinfo{journal}{Rev. Mod. Phys.} \textbf{\bibinfo{volume}{85}},
  \bibinfo{pages}{693} (\bibinfo{year}{2013}).

\bibitem[{\citenamefont{Pachucki et~al.}(2014)\citenamefont{Pachucki,
  Puchalski, and Yerokhin}}]{Pachucki:14}
\bibinfo{author}{\bibfnamefont{K.}~\bibnamefont{Pachucki}},
  \bibinfo{author}{\bibfnamefont{M.}~\bibnamefont{Puchalski}},
  \bibnamefont{and} \bibinfo{author}{\bibfnamefont{V.}~\bibnamefont{Yerokhin}},
  \bibinfo{journal}{Computer Physics Communications}
  \textbf{\bibinfo{volume}{185}}, \bibinfo{pages}{2913 }
  (\bibinfo{year}{2014}).

\bibitem[{\citenamefont{Korobov}(2002)}]{Korobov:02}
\bibinfo{author}{\bibfnamefont{V.~I.} \bibnamefont{Korobov}},
  \bibinfo{journal}{Phys. Rev. A} \textbf{\bibinfo{volume}{66}},
  \bibinfo{pages}{024501} (\bibinfo{year}{2002}).

\bibitem[{\citenamefont{Drake et~al.}(2002)\citenamefont{Drake, Cassar, and
  Nistor}}]{Drake:02}
\bibinfo{author}{\bibfnamefont{G.~W.~F.} \bibnamefont{Drake}},
  \bibinfo{author}{\bibfnamefont{M.~M.} \bibnamefont{Cassar}},
  \bibnamefont{and} \bibinfo{author}{\bibfnamefont{R.~A.}
  \bibnamefont{Nistor}}, \bibinfo{journal}{Phys. Rev. A}
  \textbf{\bibinfo{volume}{65}}, \bibinfo{pages}{054501}
  (\bibinfo{year}{2002}).

\bibitem[{\citenamefont{Pachucki and Komasa}(2009)}]{PK09}
\bibinfo{author}{\bibfnamefont{K.}~\bibnamefont{Pachucki}} \bibnamefont{and}
  \bibinfo{author}{\bibfnamefont{J.}~\bibnamefont{Komasa}},
  \bibinfo{journal}{J. Chem. Phys.} \textbf{\bibinfo{volume}{130}},
  \bibinfo{pages}{164113} (\bibinfo{year}{2009}).

\bibitem[{\citenamefont{Pachucki}(2010)}]{Pachucki:10b}
\bibinfo{author}{\bibfnamefont{K.}~\bibnamefont{Pachucki}},
  \bibinfo{journal}{Phys. Rev. A} \textbf{\bibinfo{volume}{82}},
  \bibinfo{pages}{032509} (\bibinfo{year}{2010}).

\bibitem[{\citenamefont{Pachucki and Komasa}(2014)}]{PK14}
\bibinfo{author}{\bibfnamefont{K.}~\bibnamefont{Pachucki}} \bibnamefont{and}
  \bibinfo{author}{\bibfnamefont{J.}~\bibnamefont{Komasa}},
  \bibinfo{journal}{J. Chem. Phys.} \textbf{\bibinfo{volume}{141}},
  \bibinfo{eid}{224103} (\bibinfo{year}{2014}).

\bibitem[{\citenamefont{Pachucki and Komasa}(2015)}]{PK15}
\bibinfo{author}{\bibfnamefont{K.}~\bibnamefont{Pachucki}} \bibnamefont{and}
  \bibinfo{author}{\bibfnamefont{J.}~\bibnamefont{Komasa}},
  \bibinfo{journal}{J. Chem. Phys.} \textbf{\bibinfo{volume}{143}},
  \bibinfo{eid}{034111} (\bibinfo{year}{2015}).

\bibitem[{\citenamefont{Komasa et~al.}(2011)\citenamefont{Komasa,
  Piszczatowski, {\L}ach, Przybytek, Jeziorski, and Pachucki}}]{Komasa:11}
\bibinfo{author}{\bibfnamefont{J.}~\bibnamefont{Komasa}},
  \bibinfo{author}{\bibfnamefont{K.}~\bibnamefont{Piszczatowski}},
  \bibinfo{author}{\bibfnamefont{G.}~\bibnamefont{{\L}ach}},
  \bibinfo{author}{\bibfnamefont{M.}~\bibnamefont{Przybytek}},
  \bibinfo{author}{\bibfnamefont{B.}~\bibnamefont{Jeziorski}},
  \bibnamefont{and} \bibinfo{author}{\bibfnamefont{K.}~\bibnamefont{Pachucki}},
  \bibinfo{journal}{J. Chem. Theory Comput.} \textbf{\bibinfo{volume}{7}},
  \bibinfo{pages}{3105} (\bibinfo{year}{2011}).

\bibitem[{\citenamefont{Liu et~al.}(2009)\citenamefont{Liu, Salumbides,
  Hollenstein, Koelemeij, Eikema, Ubachs, and Merkt}}]{Liu:09}
\bibinfo{author}{\bibfnamefont{J.}~\bibnamefont{Liu}},
  \bibinfo{author}{\bibfnamefont{E.~J.} \bibnamefont{Salumbides}},
  \bibinfo{author}{\bibfnamefont{U.}~\bibnamefont{Hollenstein}},
  \bibinfo{author}{\bibfnamefont{J.~C.~J.} \bibnamefont{Koelemeij}},
  \bibinfo{author}{\bibfnamefont{K.~S.~E.} \bibnamefont{Eikema}},
  \bibinfo{author}{\bibfnamefont{W.}~\bibnamefont{Ubachs}}, \bibnamefont{and}
  \bibinfo{author}{\bibfnamefont{F.}~\bibnamefont{Merkt}}, \bibinfo{journal}{J.
  Chem. Phys.} \textbf{\bibinfo{volume}{130}}, \bibinfo{eid}{174306}
  (\bibinfo{year}{2009}).

\bibitem[{\citenamefont{Stanke and Adamowicz}(2013)}]{Stanke:13}
\bibinfo{author}{\bibfnamefont{M.}~\bibnamefont{Stanke}} \bibnamefont{and}
  \bibinfo{author}{\bibfnamefont{L.}~\bibnamefont{Adamowicz}},
  \bibinfo{journal}{J. Phys. Chem. A} \textbf{\bibinfo{volume}{117}},
  \bibinfo{pages}{10129} (\bibinfo{year}{2013}).

\bibitem[{\citenamefont{Cheng et~al.}(2012)\citenamefont{Cheng, Sun, Pan, Wang,
  Liu, Campargue, and Hu}}]{Cheng:12}
\bibinfo{author}{\bibfnamefont{C.-F.} \bibnamefont{Cheng}},
  \bibinfo{author}{\bibfnamefont{Y.~R.} \bibnamefont{Sun}},
  \bibinfo{author}{\bibfnamefont{H.}~\bibnamefont{Pan}},
  \bibinfo{author}{\bibfnamefont{J.}~\bibnamefont{Wang}},
  \bibinfo{author}{\bibfnamefont{A.-W.} \bibnamefont{Liu}},
  \bibinfo{author}{\bibfnamefont{A.}~\bibnamefont{Campargue}},
  \bibnamefont{and} \bibinfo{author}{\bibfnamefont{S.-M.} \bibnamefont{Hu}},
  \bibinfo{journal}{Phys. Rev. A} \textbf{\bibinfo{volume}{85}},
  \bibinfo{pages}{024501} (\bibinfo{year}{2012}).

\bibitem[{\citenamefont{Niu et~al.}(2014)\citenamefont{Niu, Salumbides,
  Dickenson, Eikema, and Ubachs}}]{Niu:14}
\bibinfo{author}{\bibfnamefont{M.}~\bibnamefont{Niu}},
  \bibinfo{author}{\bibfnamefont{E.}~\bibnamefont{Salumbides}},
  \bibinfo{author}{\bibfnamefont{G.}~\bibnamefont{Dickenson}},
  \bibinfo{author}{\bibfnamefont{K.}~\bibnamefont{Eikema}}, \bibnamefont{and}
  \bibinfo{author}{\bibfnamefont{W.}~\bibnamefont{Ubachs}},
  \bibinfo{journal}{J. Mol. Spectrosc.} \textbf{\bibinfo{volume}{300}},
  \bibinfo{pages}{44 } (\bibinfo{year}{2014}).

\bibitem[{\citenamefont{Liu et~al.}(2010)\citenamefont{Liu, Sprecher, Jungen,
  Ubachs, and Merkt}}]{Liu:10}
\bibinfo{author}{\bibfnamefont{J.}~\bibnamefont{Liu}},
  \bibinfo{author}{\bibfnamefont{D.}~\bibnamefont{Sprecher}},
  \bibinfo{author}{\bibfnamefont{C.}~\bibnamefont{Jungen}},
  \bibinfo{author}{\bibfnamefont{W.}~\bibnamefont{Ubachs}}, \bibnamefont{and}
  \bibinfo{author}{\bibfnamefont{F.}~\bibnamefont{Merkt}}, \bibinfo{journal}{J.
  Chem. Phys.} \textbf{\bibinfo{volume}{132}}, \bibinfo{pages}{154301}
  (\bibinfo{year}{2010}).

\bibitem[{\citenamefont{Mondelain et~al.}(2016)\citenamefont{Mondelain, Kassi,
  Sala, Romanini, Gatti, and Campargue}}]{Mondelain:16}
\bibinfo{author}{\bibfnamefont{D.}~\bibnamefont{Mondelain}},
  \bibinfo{author}{\bibfnamefont{S.}~\bibnamefont{Kassi}},
  \bibinfo{author}{\bibfnamefont{T.}~\bibnamefont{Sala}},
  \bibinfo{author}{\bibfnamefont{D.}~\bibnamefont{Romanini}},
  \bibinfo{author}{\bibfnamefont{D.}~\bibnamefont{Gatti}}, \bibnamefont{and}
  \bibinfo{author}{\bibfnamefont{A.}~\bibnamefont{Campargue}},
  \bibinfo{journal}{J. Mol. Spectrosc.} \textbf{\bibinfo{volume}{326}},
  \bibinfo{pages}{5 } (\bibinfo{year}{2016}).

\end{thebibliography}


\begin{widetext}  
   
\begin{table}[!htb]
\renewcommand{\arraystretch}{1.3}
\caption{Convergence of matrix elements of relativistic operators at $R=0$ a.u. (helium atom limit).}
\label{TBL-helium}
\begin{ruledtabular}
\begin{tabular}{c@{\extracolsep{\fill}}w{6.8}w{3.8}w{2.19}}
\centt{Basis} & \centt{Direct} & \centt{Standard regularization} & \centt{\rECG(+modified regularization)} \\
\hline
\multicolumn{4}{c}{$\mathcal{E}$}\\
128      & \multicolumn{2}{w{3.8}}{-2.903\,724\,368\,357\,561} & -2.903\,724\,366\,011\,805 \\
256      & \multicolumn{2}{w{3.8}}{-2.903\,724\,376\,781\,020} & -2.903\,724\,376\,765\,067 \\
512      & \multicolumn{2}{w{3.8}}{-2.903\,724\,377\,031\,170} & -2.903\,724\,377\,030\,040 \\
1024     & \multicolumn{2}{w{3.8}}{-2.903\,724\,377\,034\,103} & -2.903\,724\,377\,034\,089 \\[1ex]
$\infty$-Slater$^a$ & & & -2.903\,724\,377\,034\,119\,598(1)\\
\multicolumn{4}{c}{$p_1^4+p_2^4$}\\
128      & 108.103\,812\,847 & 108.178\,260\,879 & 108.175\,893\,984 \\
256      & 108.149\,717\,136 & 108.176\,705\,311 & 108.176\,119\,036 \\
512      & 108.171\,069\,063 & 108.176\,261\,126 & 108.176\,133\,296 \\
1024     & 108.174\,593\,664 & 108.176\,173\,934 & 108.176\,134\,411 \\[1ex]
$\infty$-Slater$^a$ & & & 108.176\,134\,45(1) \\
\multicolumn{4}{c}{$\delta^3(r_1)+\delta^3(r_2)$}\\
128      & 3.618\,072\,922\,23 & 3.620\,855\,927\,07 & 3.620\,852\,504\,90 \\
256      & 3.619\,832\,314\,29 & 3.620\,858\,327\,71 & 3.620\,858\,263\,22 \\
512      & 3.620\,662\,493\,08 & 3.620\,858\,623\,04 & 3.620\,858\,610\,86 \\
1024     & 3.620\,798\,945\,59 & 3.620\,858\,636\,28 & 3.620\,858\,636\,16 \\[1ex]
$\infty$-Slater$^a$ & & &3.620\,858\,637\,00(1) \\
\multicolumn{4}{c}{$\delta^3(r)$}\\
128      & 0.106\,521\,423\,626 & 0.106\,345\,075\,042 & 0.106\,345\,517\,181 \\
256      & 0.106\,391\,759\,156 & 0.106\,345\,347\,318 & 0.106\,345\,416\,874 \\
512      & 0.106\,355\,477\,797 & 0.106\,345\,369\,617 & 0.106\,345\,375\,554 \\
1024     & 0.106\,348\,511\,028 & 0.106\,345\,370\,530 & 0.106\,345\,370\,708 \\[1ex]
$\infty$-Slater$^a$ & & & 0.106\,345\,370\,634(1) \\
\multicolumn{4}{c}{$p_1^i \Bigl( \frac{\delta^{ij}}{r} + \frac{r^i r^j}{r^{3}} \Bigr) p_2^j $}\\
128      & \multicolumn{2}{c}{0.278\,191\,140\,60} & 0.278\,188\,211\,08 \\
256      & \multicolumn{2}{c}{0.278\,189\,536\,41} & 0.278\,188\,961\,40 \\
512      & \multicolumn{2}{c}{0.278\,189\,388\,13} & 0.278\,189\,339\,07 \\
1024     & \multicolumn{2}{c}{0.278\,189\,381\,79} & 0.278\,189\,380\,36 \\
$\infty$-Slater$^a$ & & & 0.278\,189\,381\,08(1)  \\
\multicolumn{4}{c}{$\mathcal{E}_\mathrm{rel}$}\\
128      & -1.950\,913\,941\,68 & -1.952\,030\,893\,54 & -1.951\,742\,928\,89 \\
256      & -1.951\,531\,235\,51 & -1.951\,827\,248\,21 & -1.951\,753\,660\,41 \\
512      & -1.951\,706\,049\,93 & -1.951\,770\,653\,11 & -1.951\,754\,696\,40 \\
1024     & -1.951\,739\,830\,39 & -1.951\,759\,709\,00 & -1.951\,754\,765\,23 \\[1ex]
$\infty$-Slater$^a$ & & & -1.951\,754\,768(1) \\
\end{tabular}                            
\end{ruledtabular}     
\flushleft
$^a$ The reference values were evaluated with the atomic ECE basis functions 
$\phi=\exp(-\alpha\,r_1-\beta\,r_2-\gamma\,r)$.
\end{table}                              

\begin{table}[!htb]
\renewcommand{\arraystretch}{1.3}
\caption{Convergence of matrix elements of relativistic operators at the equilibrium distance $R=1.4$ a.u.}
\label{TBL3}
\begin{ruledtabular}
\begin{tabular}{c@{\extracolsep{\fill}}w{6.8}w{3.8}w{2.19}}
\centt{Basis} & \centt{Direct} & \centt{Standard regularization} & \centt{\rECG(+modified regularization)} \\
\hline
\multicolumn{4}{c}{$\mathcal{E}$}\\
128      & \multicolumn{2}{c}{-1.174\,475\,621\,659\,802} & -1.174\,475\,640\,130\,736 \\
256      & \multicolumn{2}{c}{-1.174\,475\,711\,731\,700} & -1.174\,475\,711\,200\,533 \\
512      & \multicolumn{2}{c}{-1.174\,475\,714\,117\,150} & -1.174\,475\,714\,015\,654 \\
1024     & \multicolumn{2}{c}{-1.174\,475\,714\,217\,171} & -1.174\,475\,714\,203\,071 \\
$\infty$-JC$^a$ & \multicolumn{2}{c}{.}  &  -1.174\,475\,714\,220\,443\,4(5)
 \\
\multicolumn{4}{c}{$p_1^4+p_2^4$}\\
128      & 13.214\,563 & 13.238\,771\,4 & 13.237\,780\,507 \\
256      & 13.231\,849 & 13.238\,308\,1 & 13.237\,929\,826 \\
512      & 13.235\,568 & 13.238\,045\,9 & 13.237\,954\,576 \\
1024     & 13.237\,266 & 13.237\,981\,4 & 13.237\,956\,021 \\
$\infty$ & 13.238\,7(10)    & 13.237\,960(16)  & 13.237\,956\,18(7) \\
\multicolumn{4}{c}{$\sum_{a,X}\delta^3(r_{aX})$}\\
128      & 0.917\,550\,7 & 0.919\,331\,278\,50 & 0.919\,331\,321\,06 \\
256      & 0.918\,878\,9 & 0.919\,335\,927\,94 & 0.919\,335\,741\,10 \\
512      & 0.919\,153\,5 & 0.919\,336\,172\,60 & 0.919\,336\,127\,42 \\
1024     & 0.919\,285\,5 & 0.919\,336\,210\,05 & 0.919\,336\,191\,74 \\
$\infty$ & 0.919\,37(7)  & 0.919\,336\,211\,2(18)& 0.919\,336\,206(7) \\
JC$^b$   &               &                       & 0.919\,336\,210\,2 \\
\multicolumn{4} {c}{$\delta^3(r)$}\\
128      & 0.016\,742\,915 & 0.016\,742\,915\,953 & 0.016\,743\,529\,776 \\
256      & 0.016\,771\,639 & 0.016\,743\,229\,495 & 0.016\,743\,316\,316 \\
512      & 0.016\,750\,461 & 0.016\,743\,274\,514 & 0.016\,743\,287\,361 \\
1024     & 0.016\,745\,258 & 0.016\,743\,277\,598 & 0.016\,743\,278\,963 \\
$\infty$ & 0.016\,743\,5(4)& 0.016\,743\,278\,1(7)& 0.016\,743\,278\,3(5)\\
JC$^b$   &                 &                      & 0.016\,743\,278\,80 \\
\multicolumn{4}{c}{$p_1^i \Bigl( \frac{\delta^{ij}}{r} + \frac{r^i r^j}{r^{3}} \Bigr) p_2^j$}\\
128      & \multicolumn{2}{c}{0.095\,271\,222\,30} & 0.095\,266\,566\,34 \\
256      & \multicolumn{2}{c}{0.095\,269\,308\,16} & 0.095\,268\,588\,45 \\
512      & \multicolumn{2}{c}{0.095\,269\,010\,11} & 0.095\,268\,883\,81\\
1024     & \multicolumn{2}{c}{0.095\,268\,989\,79} & 0.095\,268\,976\,83 \\
$\infty$ & \multicolumn{2}{c}{0.095\,268\,986(6)}   & 0.095\,268\,987(4)\\
\multicolumn{4}{c}{$\mathcal{E}_\mathrm{rel}$}\\
128      & -0.205\,342\,30 & -0.205\,800\,42 & -0.205\,672\,234\,22 \\
256      & -0.205\,554\,44 & -0.205\,733\,26 & -0.205\,685\,637\,69 \\                        
512      & -0.205\,654\,50 & -0.205\,699\,82 & -0.205\,688\,363\,37 \\                        
1024     & -0.205\,675\,65 & -0.205\,691\,67 & -0.205\,688\,516\,45 \\
$\infty$ & -0.205\,682(8)  & -0.205\,689(2)  & -0.205\,688\,526(7) \\
\end{tabular}                            
\end{ruledtabular}   
\flushleft
$^a$ Evaluated with James-Coolidge wave function, Ref.~\cite{Pachucki:10b}.\\
$^b$ Evaluated in this work.
\end{table}                              

\begin{table}[!htb]
\small 
\renewcommand{\arraystretch}{1.1}
\caption{The electronic energy $\mathcal{E}$, expectation values of individual relativistic operators, 
and the relativistic correction of Eq.~(\ref{HBPr}) evaluated with 1024-term \rECG basis for H$_2$ 
(all entries in a.u.).}
\label{TBL4}
\begin{ruledtabular}
\begin{tabular}{w{2.4}w{2.17}w{3.10}w{1.12}w{1.15}w{2.14}w{2.11}}
\cent{R} & \cent{\cal E} & \cent{p_1^4+p_2^4} & \cent{\sum_{a,X}\delta^3(r_{aX})} & \cent{\delta^3(r)} & 
\cent{p_1^i \Bigl( \frac{\delta^{ij}}{r} + \frac{r^i r^j}{r^3} \Bigr) p_2^j } & \cent{{\cal E}_\mathrm{rel}}  \\[2ex]
\hline
0.0      & \cent{\infty}            &108.176\,134\,41 & 7.241\,717\,273\,9 & 0.106\,345\,370\,636 &  0.278\,189\,381\,06 & -1.951\,754\,765  \\
0.05     & 17.104\,840\,595\,733\,8 & 97.883\,437\,11 & 6.489\,043\,177\,5 & 0.105\,003\,823\,908 &  0.276\,266\,361\,42 & -1.850\,717\,847 \\
0.1      &  7.127\,216\,731\,179\,9 & 87.359\,441\,20 & 5.770\,597\,148\,6 & 0.101\,570\,564\,053 &  0.271\,273\,759\,54 & -1.672\,040\,901 \\
0.2      &  2.197\,803\,295\,242\,6 & 69.106\,095\,40 & 4.568\,931\,840\,2 & 0.091\,368\,192\,666 &  0.255\,825\,956\,08 & -1.302\,271\,866 \\ 
0.4      & -0.120\,230\,341\,173\,2 & 44.926\,748\,48 & 3.007\,974\,262\,0 & 0.068\,873\,279\,242 &  0.218\,115\,094\,21 & -0.783\,614\,393 \\
0.6      & -0.769\,635\,429\,474\,0 & 31.455\,544\,41 & 2.135\,554\,220\,2 & 0.050\,763\,240\,175 &  0.183\,036\,489\,08 & -0.509\,463\,149 \\
0.8      & -1.020\,056\,666\,340\,7 & 23.551\,412\,69 & 1.616\,559\,002\,3 & 0.037\,667\,878\,063 &  0.153\,865\,267\,48 & -0.363\,237\,148 \\
1.0       & -1.124\,539\,719\,525\,6 & 18.631\,892\,59 & 1.288\,195\,811\,0 & 0.028\,345\,276\,112 &  0.130\,210\,595\,68 & -0.281\,549\,312 \\
1.1      & -1.150\,057\,367\,720\,2 & 16.862\,669\,10 & 1.168\,538\,437\,1 & 0.024\,725\,634\,292 &  0.120\,121\,309\,61 & -0.254\,680\,537 \\
1.2      & -1.164\,935\,243\,421\,7 & 15.417\,026\,26 & 1.069\,918\,912\,8 & 0.021\,643\,994\,354 &  0.111\,012\,509\,31 & -0.234\,013\,225 \\
1.3      & -1.172\,347\,149\,015\,1 & 14.225\,963\,31 & 0.987\,949\,432\,3 & 0.019\,008\,309\,934 &  0.102\,764\,094\,19 & -0.218\,043\,755 \\
1.4      & -1.174\,475\,714\,202\,3 & 13.237\,956\,02 & 0.919\,336\,191\,7 & 0.016\,743\,278\,963 &  0.095\,268\,976\,86 & -0.205\,688\,516 \\
1.4011   & -1.174\,475\,931\,376\,0 & 13.228\,062\,27 & 0.918\,645\,775\,7 & 0.016\,720\,183\,220 &  0.095\,190\,366\,90 & -0.205\,569\,553 \\
1.45     & -1.174\,057\,071\,449\,9 & 12.807\,526\,10 & 0.889\,230\,483\,6 & 0.015\,730\,076\,395 &  0.091\,773\,957\,00 & -0.200\,610\,272 \\
1.5      & -1.172\,855\,079\,551\,8 & 12.414\,016\,28 & 0.861\,572\,931\,7 & 0.014\,787\,413\,515 &  0.088\,432\,824\,23 & -0.196\,156\,821 \\
1.6      & -1.168\,583\,373\,346\,2 & 11.724\,276\,68 & 0.812\,728\,747\,6 & 0.013\,090\,471\,717 &  0.082\,173\,146\,96 & -0.188\,864\,898 \\
1.7      & -1.162\,458\,726\,874\,9 & 11.145\,591\,08 & 0.771\,298\,695\,7 & 0.011\,611\,319\,478 &  0.076\,418\,016\,59 & -0.183\,376\,700 \\
1.8      & -1.155\,068\,737\,586\,8 & 10.659\,827\,14 & 0.736\,097\,157\,4 & 0.010\,316\,180\,045 &  0.071\,104\,900\,35 & -0.179\,362\,896 \\
1.9      & -1.146\,850\,697\,001\,6 & 10.252\,633\,62 & 0.706\,180\,601\,1 & 0.009\,177\,235\,715 &  0.066\,179\,438\,68 & -0.176\,571\,891 \\
2.0      & -1.138\,132\,957\,102\,2 &  9.912\,536\,44 & 0.680\,790\,743\,4 & 0.008\,171\,495\,990 &  0.061\,594\,536\,96 & -0.174\,809\,213 \\
2.1      & -1.129\,163\,836\,066\,7 &  9.630\,266\,86 & 0.659\,312\,181\,5 & 0.007\,279\,910\,924 &  0.057\,309\,396\,71 & -0.173\,922\,389 \\
2.2      & -1.120\,132\,116\,815\,1 &  9.398\,255\,83 & 0.641\,240\,366\,5 & 0.006\,486\,644\,310 &  0.053\,288\,924\,14 & -0.173\,790\,035 \\
2.3      & -1.111\,181\,765\,169\,5 &  9.210\,246\,40 & 0.626\,157\,014\,6 & 0.005\,778\,509\,183 &  0.049\,503\,041\,98 & -0.174\,313\,461 \\
2.4      & -1.102\,422\,605\,975\,9 &  9.060\,993\,78 & 0.613\,710\,992\,0 & 0.005\,144\,501\,580 &  0.045\,926\,280\,03 & -0.175\,410\,463 \\
2.5      & -1.093\,938\,129\,920\,1 &  8.946\,029\,98 & 0.603\,603\,282\,8 & 0.004\,575\,430\,210 &  0.042\,537\,333\,71 & -0.177\,010\,457 \\
2.6      & -1.085\,791\,237\,362\,5 &  8.861\,476\,93 & 0.595\,575\,000\,5 & 0.004\,063\,603\,931 &  0.039\,318\,826\,34 & -0.179\,050\,818 \\
2.7      & -1.078\,028\,484\,147\,9 &  8.803\,898\,42 & 0.589\,397\,852\,6 & 0.003\,602\,589\,090 &  0.036\,256\,887\,42 & -0.181\,473\,897 \\
2.8      & -1.070\,683\,233\,449\,8 &  8.770\,183\,99 & 0.584\,866\,570\,4 & 0.003\,186\,984\,950 &  0.033\,341\,021\,16 & -0.184\,225\,040 \\
2.9      & -1.063\,778\,008\,771\,7 &  8.757\,458\,14 & 0.581\,792\,947\,1 & 0.002\,812\,250\,937 &  0.030\,563\,703\,83 & -0.187\,250\,948 \\
3.0      & -1.057\,326\,268\,838\,3 &  8.763\,015\,72 & 0.580\,001\,450\,5 & 0.002\,474\,548\,852 &  0.027\,920\,059\,06 & -0.190\,498\,823 \\
3.2      & -1.045\,799\,661\,390\,2 &  8.818\,766\,65 & 0.579\,608\,528\,2 & 0.001\,897\,578\,330 &  0.023\,025\,370\,71 & -0.197\,450\,152 \\
3.4      & -1.036\,075\,395\,153\,1 &  8.918\,013\,44 & 0.582\,448\,331\,7 & 0.001\,434\,600\,771 &  0.018\,655\,309\,93 & -0.204\,664\,704 \\
3.6      & -1.028\,046\,308\,339\,0 &  9.043\,109\,19 & 0.587\,399\,895\,4 & 0.001\,068\,345\,909 &  0.014\,820\,897\,94 & -0.211\,757\,192 \\
3.8      & -1.021\,549\,795\,379\,5 &  9.178\,853\,16 & 0.593\,491\,357\,9 & 0.000\,783\,739\,899 &  0.011\,529\,591\,76 & -0.218\,405\,205 \\
4.0      & -1.016\,390\,252\,917\,8 &  9.313\,279\,23 & 0.599\,942\,199\,5 & 0.000\,566\,901\,578 &  0.008\,773\,121\,69 & -0.224\,378\,488 \\
4.2      & -1.012\,359\,959\,653\,3 &  9.438\,064\,55 & 0.606\,187\,454\,4 & 0.000\,404\,933\,730 &  0.006\,522\,438\,89 & -0.229\,550\,124 \\
4.4      & -1.009\,256\,516\,218\,8 &  9.548\,395\,99 & 0.611\,872\,426\,5 & 0.000\,286\,163\,107 &  0.004\,729\,595\,59 & -0.233\,888\,328 \\
4.6      & -1.006\,895\,223\,788\,3 &  9.642\,347\,77 & 0.616\,819\,321\,1 & 0.000\,200\,467\,812 &  0.003\,334\,249\,31 & -0.237\,433\,284 \\
4.8      & -1.005\,116\,006\,012\,2 &  9.720\,046\,26 & 0.620\,980\,427\,9 & 0.000\,139\,467\,088 &  0.002\,271\,426\,58 & -0.240\,269\,571 \\
5.0      & -1.003\,785\,658\,541\,8 &  9.782\,862\,62 & 0.624\,391\,465\,4 & 0.000\,096\,514\,570 &  0.001\,478\,031\,56 & -0.242\,501\,813 \\
5.2      & -1.002\,796\,816\,280\,7 &  9.832\,762\,96 & 0.627\,132\,969\,7 & 0.000\,066\,524\,898 &  0.000\,897\,184\,78 & -0.244\,236\,803 \\
5.4      & -1.002\,065\,057\,189\,4 &  9.871\,870\,41 & 0.629\,303\,290\,7 & 0.000\,045\,719\,642 &  0.000\,480\,298\,37 & -0.245\,573\,020 \\
5.6      & -1.001\,525\,251\,817\,8 &  9.902\,201\,97 & 0.631\,001\,580\,7 & 0.000\,031\,354\,765 &  0.000\,187\,505\,16 & -0.246\,595\,529 \\
5.8      & -1.001\,127\,880\,827\,6 &  9.925\,541\,76 & 0.632\,318\,772\,1 & 0.000\,021\,470\,798 & -0.000\,012\,912\,40 & -0.247\,374\,807 \\
6.0      & -1.000\,835\,707\,602\,8 &  9.943\,392\,93 & 0.633\,333\,420\,5 & 0.000\,014\,686\,948 & -0.000\,145\,606\,65 & -0.247\,967\,362 \\
6.5      & -1.000\,400\,547\,946\,1 &  9.971\,478\,81 & 0.634\,946\,087\,8 & 0.000\,005\,668\,125 & -0.000\,293\,186\,06 & -0.248\,899\,471 \\
7.0      & -1.000\,197\,914\,426\,6 &  9.985\,482\,43 & 0.635\,760\,762\,7 & 0.000\,002\,183\,134 & -0.000\,307\,901\,17 & -0.249\,373\,815 \\ 
7.5      & -1.000\,102\,106\,038\,0 &  9.992\,439\,90 & 0.636\,169\,798\,9 & 0.000\,000\,839\,980 & -0.000\,274\,910\,15 & -0.249\,621\,706 \\ 
8.0      & -1.000\,055\,604\,611\,0 &  9.995\,922\,78 & 0.636\,376\,259\,1 & 0.000\,000\,323\,028 & -0.000\,230\,376\,86 & -0.249\,756\,677 \\ 
8.5      & -1.000\,032\,171\,701\,7 &  9.997\,700\,01 & 0.636\,482\,218\,0 & 0.000\,000\,124\,123 & -0.000\,188\,104\,44 & -0.249\,834\,155 \\ 
9.0      & -1.000\,019\,781\,690\,9 &  9.998\,632\,62 & 0.636\,537\,991\,1 & 0.000\,000\,047\,661 & -0.000\,152\,289\,86 & -0.249\,881\,211 \\ 
9.5      & -1.000\,012\,855\,993\,3 &  9.999\,142\,48 & 0.636\,568\,497\,4 & 0.000\,000\,018\,282 & -0.000\,123\,345\,87 & -0.249\,911\,507 \\ 
10.0     & -1.000\,008\,755\,693\,5 &  9.999\,434\,61 & 0.636\,586\,008\,8 & 0.000\,000\,007\,007 & -0.000\,100\,390\,54 & -0.249\,932\,046 \\
\infty   & -1.0                     & 10.0            & 0.636\,619\,772\,3 & 0.0                  &  0.0                 & -0.25 \\
\end{tabular}                            
\end{ruledtabular}     
\end{table}                              

\end{widetext}

\begin{table}[!htb]
\renewcommand{\arraystretch}{1.3}
\caption{Contributions to the dissociation energy $D_0$ and two selected most accurate
experimental transitions in H$_2$ (in $\icm$). There are additional $10^{-3}$ relative 
uncertainties on $E^{(2)}$, $E^{(3)}$, and $E^{(4)}$ terms due to
the BO approximation, which are included in the final result only.}
\label{TH2BO}
\begin{ruledtabular}
\begin{tabular}{l@{\extracolsep{\fill}}w{6.12}w{6.9}w{5.10}}
\centt{Contrib.} & \cent{D_0} & \cent{S_3(3)} & \cent{Q_1(0)} \\
\hline
$E$     	& 36\,118.797\,746\,12(5) & 12\,559.749\,919(1)  & 4\,161.164\,070\,3(1)   \\
$E^{(2)}$	&      -0.533\,121(1)^a   &       0.065\,366	   &      0.023\,397	    \\
$E^{(3)}$	&      -0.194\,8(2)       &      -0.065\,73(6)   &     -0.021\,29(2)	    \\
$E^{(4)}$	&      -0.002\,067(6)     &      -0.000\,599     &     -0.000\,192	    \\
$E^{(5)}$	&       0.000\,12(6)      & 	    0.000\,037(19) &      0.000\,012(6)   \\
$E_\mathrm{FS}$& -0.000\,031        & 	   -0.000\,010     &     -0.000\,003      \\
Total	    & 36\,118.067\,8(6)       & 12\,559.748\,98(8)   & 4\,161.165\,99(3)    \\
Exp.	    & 36\,118.069\,62(37)^b   & 12\,559.749\,52(5)^c & 4\,161.166\,36(15)^d \\
Diff.	    &       0.001\,8		      &       0.000\,54      &      0.000\,37       \\
\end{tabular}                            
\end{ruledtabular}   
\flushleft
$^a$ For comparison, Wolniewicz\citep{Wolniewicz:93} obtained $-0.5330$, Piszczatowski et al.\citep{PLPKPJ09} $-0.5319$,
and Stanke et al.\citep{Stanke:13} $-0.5691\,\icm$ (the latter value comes from nonadiabatic calculations).\\
$^b$ Ref.~\cite{Liu:09}; $^c$ Ref.~\cite{Cheng:12}; $^d$ Ref.~\cite{Niu:14}.
\end{table}

\begin{table}[!htb]
\renewcommand{\arraystretch}{1.3}
\caption{Contributions to the dissociation energy $D_0$ and two selected most accurate
experimental transitions in D$_2$ (in $\icm$). There are additional $5\cdot 10^{-4}$ relative 
uncertainties on $E^{(2)}$, $E^{(3)}$, and $E^{(4)}$ terms due to
the BO approximation, which are included in the final result only.}
\label{TD2BO}
\begin{ruledtabular}
\begin{tabular}{l@{\extracolsep{\fill}}w{6.9}w{6.9}w{6.9}}
\centt{Contrib.} & \cent{D_0} & \cent{S_2(2)} & \cent{Q_1(0)} \\
\hline
$E$               & 36\,749.090\,98(8)  & 6\,241.120\,96(30)  & 2\,993.614\,88(15)     \\
$E^{(2)}$					&      -0.529\,170(1)	  &      0.040\,057     &      0.017\,677      \\
$E^{(3)}$					&      -0.198\,2(2)     &     -0.033\,15(3)    &     -0.015\,39(2)	     \\
$E^{(4)}$					&      -0.002\,096(6)   &     -0.000\,299     &     -0.000\,139   	 \\
$E^{(5)}$					&       0.000\,12(6)  &      0.000\,019(10) &      0.000\,009(5)      \\
$E_\mathrm{FS} $	&      -0.000\,204		  &     -0.000\,032     &     -0.000\,015      \\
Total			        & 36\,748.361\,4(4)	      & 6\,241.127\,55(30) & 2\,993.617\,02(15)   \\
Exp.              & 36\,748.362\,86(68)^a & 6\,241.127\,64(2)^b & 2\,993.617\,06(15)^c \\
Diff.			        &       0.001\,46       &      0.000\,09      &      0.000\,04	     \\
\end{tabular}                            
\end{ruledtabular}   
\flushleft
$^a$ Ref.~\cite{Liu:10}; $^b$ Ref.~\cite{Mondelain:16}; $^c$ Ref.~\cite{Niu:14}.
\end{table}                              

\end{document}